# The Ubiquity of Small-World Networks

Qawi K. Telesford[1*], Karen E. Joyce[1], Satoru Hayasaka[2,3], Jonathan H. Burdette[3], Paul J. Laurienti[3]

[1] School of Biomedical Engineering and Sciences, Virginia Tech-Wake Forest University, Winston-Salem, NC USA
[2] Department of Biostatistical Sciences, Wake Forest University School of Medicine, Winston-Salem, North Carolina, USA
[3] Department of Radiology, Wake Forest University School of Medicine, Winston-Salem, North Carolina, USA

**Abstract**

Small-world networks by Watts and Strogatz are a class of networks that are highly clustered, like regular lattices, yet have small characteristic path lengths, like random graphs. These characteristics result in networks with unique properties of regional specialization with efficient information transfer. Social networks are intuitive examples of this organization with cliques or clusters of friends being interconnected, but each person is really only 5-6 people away from anyone else. While this qualitative definition has prevailed in network science theory, in application, the standard quantitative application is to compare path length (a surrogate measure of distributed processing) and clustering (a surrogate measure of regional specialization) to an equivalent random network. It is demonstrated here that comparing network clustering to that of a random network can result in aberrant findings and networks once thought to exhibit small-world properties may not. We propose a new small-world metric, $\omega$ (omega), which compares network clustering to an equivalent *lattice* network and path length to a *random* network, as Watts and Strogatz originally described. Example networks are presented that would be interpreted as small-world when clustering is compared to a random network but are not small-world according to $\omega$. These findings have significant implications in network science as small-world networks have unique topological properties, and it is critical to accurately distinguish them from networks without simultaneous high clustering and low path length.

**Introduction**

The discovery of small-world networks has revolutionized research in network science. In their 1998 landmark paper, Watts and Strogatz described networks that are "highly clustered, like regular lattices, yet have small characteristic path lengths, like random graphs,"(Watts and Strogatz 1998). In other words, small-world networks have the unique ability to have specialized nodes or regions within a network (e.g., a computer network with a group of machines dedicated to a certain task) while simultaneously exhibiting shared or distributed processing across all of the communicating nodes within a network (e.g., all computers sharing the work load). Since that original paper, numerous networks have been described as exhibiting small-world properties, including systems as diverse as the internet, social groups, and biochemical pathways. Given the unique processing or information transfer capabilities of small-world networks, it is vital to determine if this is a universal property of naturally occurring networks or if small-world properties are restricted to specialized networks. An overly liberal definition of small-worldness might miss the specific benefits of these networks, high clustering and low path length, and obscure them with networks more closely associated with regular lattices and random networks.

The current accepted definition of a small-world network is that it has clustering comparable to a regular lattice and path length comparable to a random network. However, in practice, networks are typically defined as small-world by comparing clustering and path length to that of a comparable random network(Humphries et al. 2006). Unfortunately, this means that networks with very low clustering can be, and indeed are, defined as small-world. It is proposed here that such a method is unable to distinguish true small-world networks from those that are more closely aligned with either random or lattice structures, and overestimates the occurrence of small-world networks. Networks that are more similar to random or lattice structures are interesting in their own right, but they do not behave like small-world networks. Having a metric to accurately characterize networks as small-world, random, or lattice, or at least tendencies toward one of these types of networks is an extremely important factor in the study of network science. In this paper, a new metric is defined that quantifies small-world properties and places the network in question along a continuum from lattice to small-world to random. This new metric clearly demonstrates that small-world networks are not as ubiquitous as reported and suggests that many systems originally thought to have small-world processing capabilities may in fact not.

*Identifying Small-World Networks*

Small-world networks are distinguished from other networks by two specific properties, the first being high clustering (C) among nodes. Mathematically, $C$ is the proportion of edges $e_i$ that exist between the neighbors of a particular node ($i$) relative to the total number of possible edges between neighbors(Bullmore and Sporns 2009). The equation for $C$ at an individual node of degree $k_i$ is:

$$C_i = \frac{2e_i}{k_i(k_i - 1)} \qquad [1]$$

The overall clustering in a network can be determined by averaging the clustering across all individual nodes. High clustering supports specialization as local collections of strongly interconnected nodes readily share information or resources. Conceptually, clustering is quite straightforward to comprehend. In a real world analogy, clustering represents the probability that one's friends are also friends of each other.

Small-world networks also have short path lengths ($L$) as is commonly observed in random networks. Path length is a measure of the distance between nodes in the network, calculated as the mean of the shortest geodesic distances between all possible node pairs:

$$L = \frac{1}{N(N-1)} \sum_{\text{if } n, i \neq j} d_{ij} \qquad [2]$$

where $d_{ij}$ is the shortest geodesic distance between nodes $i$ and $j$.

Small values of *L* ensure that information or resources easily spreads throughout the network. This property makes distributed information processing possible on technological networks and supports the six degrees of separation often reported in social networks.

Decades of research has established that short path length is a characteristic of random graphs while high clustering is a property of lattice networks. Watts and Strogatz developed a network model (WS model) that resulted in the first ever networks with clustering close to that of a lattice and path lengths similar to random networks. The WS model demonstrates that random rewiring of a small percentage of the edges in a lattice results in a precipitous fall in the path length, but only trivial reductions in the clustering (Figure 1). Across this rewiring probability, there is a range where the discrepancy between clustering and path length is very large, and it is in this area that the benefits of small-world networks are realized.

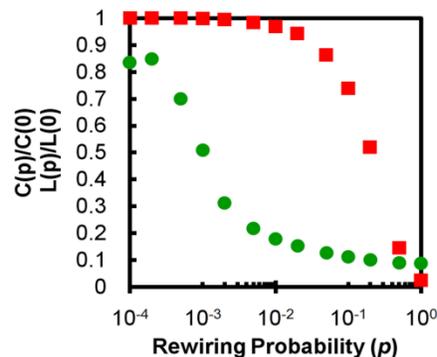

**Figure 1.** Watts and Strogatz small-world model. A simulated 1000-node lattice with average degree $k=10$ was rewired at varying probabilities $p$, ranging from 0 to 1. At small values $p$, network small-worldness is seen in a network with simultaneous high clustering (squares) and low path length (circles).

In the investigation of small-world properties, it has become common to compare both clustering and path length of a network of interest to those same metrics from an equivalent random network. In 2006, Humphries and colleagues introduced a quantitative metric, small-world coefficient $\sigma$, that utilizes a ratio of network clustering and path length compared to its random network equivalent(Humphries et al. 2006). This metric has since gained considerable popularity(Bassett et al. 2008; Liu et al. 2008; Guye et al. 2010), particularly among neuroscientists (see(Bullmore and Sporns 2009)), and has been more extensively evaluated by Humphries and Gurney(Humphries and Gurney 2008). In this quantitative approach, $C$ and $L$ are measured against those of their equivalent derived random networks ($C_{rand}$ and $L_{rand}$, respectively) to generate the ratios $\gamma = C/C_{rand}$ and $\lambda = L/L_{rand}$. These ratios are then used to calculate the small-coefficient as:

$$\sigma = \frac{C/C_{rand}}{L/L_{rand}} = \frac{\gamma}{\lambda} \qquad [3]$$

The conditions that must be met for a network to be classified as small-world are $C >> C_{rand}$ and $L \approx L_{rand}$, which results in $\sigma > 1$. The small-world coefficient has been used to ascribe small-world properties to numerous networks ranging from the power grid to the actors network(Humphries and Gurney 2008).

Comparing path length to that of a random network makes sense—the path length of a small-world network should be short like a random network. However, the comparison of clustering to an equivalent random network does not properly capture small-world behavior, since clustering in a small-world network more closely mimics that of a lattice network. Furthermore, it is generally accepted that clustering in the original network is much greater than that of a random network. But how much greater must this clustering be to resemble that of a lattice? If a random network has $C_{rand}$ of 0.001, on average, only 1 out of every 1000 possible connections are present between the neighbors of a node. If the original network has $C$ 5-10 times greater than $C_{rand}$, then there still exists only 5-10 links out of the 1000 possible connections present. The level of clustering seen in such a network is much lower than what is typically considered for a lattice. Nonetheless, the magnitude difference between $C$ and $C_{rand}$ shows that the original network is not completely a random network. However, given the low clustering, the extent to which this network is considered small-world is also unclear. It is interesting to note that even the earliest evaluation of small-world properties in real networks compared the clustering to a random network(Watts and Strogatz 1998).

As it turns out, a major issue with $\sigma$ is that the clustering coefficient of the equivalent random network greatly influences the small-world coefficient. In the small-world coefficient equation (3), $\sigma$ utilizes the relationship between $C$ and $C_{rand}$ to determine the value of $\gamma$. Because clustering in a random network is typically extremely low (see (Watts and Strogatz 1998; Humphries and Gurney 2008)) the value of $\gamma$ can be unduly influenced by only small changes in $C_{rand}$. Consider two networks $A$ and $B$ with similar path lengths, yet disparate clustering coefficients of 0.5 and 0.05, respectively. If the clustering coefficients of the equivalent random networks are both 0.01, then network $A$ clearly has greater small-world properties. However, if the clustering of the random networks for $A$ and $B$ are 0.01 and 0.001, respectively, then the two networks will have similar values of $\sigma$. Interpretation of $\sigma$ suggests that both networks have the same small-world characteristics despite the fact that network $B$ has considerably lower clustering. While it is true that relative to the comparable random networks, these two networks have the same values of $\sigma$, they clearly have different levels of clustering, and network $A$ appears to be more closely aligned with a lattice than network $B$. This finding occurs because $C_{rand}$ is in the denominator of the equation for $\gamma$; thus, small changes in $C_{rand}$ ultimately drive the value of $\sigma$.

The comparison of clustering to a random network presents several limitations to the use of $\sigma$. For example, values of $\sigma$ range from 0 to $\infty$ and are dependent on the size of the network in question(Humphries and Gurney 2008). Larger networks with comparable clustering and path length tend to have higher values of $\sigma$ than their smaller counterparts. Finally, it may be valuable to know if a network has properties that tend to be more lattice-like or random-like. It is not possible to determine these properties if clustering and path length are compared to a random equivalent. It is of great importance to determine if networks exhibit specific behaviors, such as specialization (lattices) or ability to effectively transmit information (random networks). The current work explores a new metric to quantify small-world properties, $\omega$, that addresses each of the limitations described above and is more in keeping with the original description of small-world networks as defined by Watts and Strogatz.

*Novel Small-World Measurement: $\omega$*

Given a graph with characteristic path length, *L,* and clustering, *C*, the small-world measurement, $\omega$, is defined by comparing the clustering of the network to that of an equivalent lattice network, $C_{latt}$, and path length to that of an equivalent random network, $L_{rand}$; the relationship is simply the difference of two ratios defined as:

$$\omega = \frac{L_{rand}}{L} - \frac{C}{C_{latt}} \qquad [4]$$

In using the clustering of an equivalent lattice network rather than a random network, this metric is less susceptible to the fluctuations seen with $C_{rand}$. Moreover, values of $\omega$ are restricted to the interval -1 to 1 regardless of network size. Values close to zero are considered small world: near zero, $L \approx L_{rand}$ and $C \approx C_{latt}$. Positive values indicate a graph with more random characteristics: $L \approx L_{rand}$, and $C << C_{latt}$. Negative values indicate a graph with more regular, or lattice-like, characteristics: $L >> L_{rand}$, and $C \approx C_{latt}$.

**Methods**

*Well-known networks datasets*

Biological, social, and technological networks were obtained from various sources. All networks in this study were analyzed asbinary matrices with unweighted and undirected edges. For disconnected graphs, network analysis was done on the largest component of the network. The e-mail(Guimerà et al. 2003), and *C. elegans*metabolic network(Duch and Arenas 2005) were obtained from Alex Arenas' network datasets (http://deim.urv.cat/~aarenas/data/welcome.htm). The karate(Zachary 1977), word adjacency(Newman 2006), football(Girvan and Newman 2002), and internet networks were obtained from Mark Newman's network data sets (http://www-personal.umich.edu/~mejn/netdata/). The internet network is from unpublished data by the University of Oregon Route Views Project (http://routeviews.org/).

*Brain imaging data collection*

Brain imaging data was collected from 11 healthy older adults as part of a separate study evaluating an exercise program(Joyce et al. 2010). All data reported here are from post-treatment scans with subjects either in the control or treatment group.Whole-brain functional connectivity was evaluated using graph theory methods on a voxel-by-voxel basis(Eguíluz et al. 2005; Fox et al. 2005; van den Heuvel et al. 2008; Hayasaka and Laurienti 2010).A correlation matrix was produced by computing the Pearson correlation between all voxel pairs within fMRI time series. A threshold was applied to the correlation matrix, whereby voxel pairs above the threshold were considered connected and assigned a value of 1, and voxel pairs below the threshold were considered not connected and assigned a value of 0. This binary matrix produces an undirected, unweighted adjacency matrix ($A_{ij}$) representing the whole-brain functional connectivity for each subject.Mean graph metrics for both groups were compared for degree ($k$), clustering coefficient ($C$) and minimum path length ($L$). Scanning protocol and network analysis specifics can be found in the supplemental materials.

*Random and lattice network construction*

To calculate $\sigma$ for a given network, an equivalent random graph was created by assigning an edge to a node pair with uniform probability while maintaining the degree distribution of the original graph(Maslov and Sneppen 2002). In this study, edges for were rewired at random an average of ten times for the entire network. Network randomization was performed in 50 networks with the clustering coefficient and path length calculated for each network. The mean for both graph metrics was calculated and served as the $C_{rand}$ and $L_{rand}$ of the equivalent random network.It is important to note that $\omega$ is only valid if the comparable random network preserves the degree distribution of the original network.

The lattice network was generated using a modified version of the "latticization" algorithm (Sporns and Zwi 2004) found in the brain connectivity toolbox (http://www.brain-connectivity-toolbox.net)(Rubinov and Sporns 2010). This algorithmwas previously used to scale clustering and path length between random and lattice networks(Sporns and Zwi 2004), and to evaluate the efficiency of networks over varying cost(Achard and Bullmore 2007).The procedure is based on a Markov-chain algorithm that maintains node degree and swaps edges with uniform probability; however, swaps are only carried out if the resulting matrix has entries that are closer to the main diagonal. To optimize the clustering coefficient of the lattice network, the latticization procedure is performed over several user-defined repetitions. Storing the initial adjacency matrix and its clustering coefficient, the latticization procedure is performed on the matrix. If the clustering coefficient of the resulting matrix is lower, the initial matrix is kept and latticization is performed again on the same matrix; if the clustering coefficient is higher, then the initial adjacency matrix is replaced. This latticization process is repeated until clustering is maximized. This process results in a highly clustered network with long path length approximating a lattice topology. To decrease the processing time in larger networks, a "sliding window" procedure was developed. Smaller sections of the matrix are sampled along the main diagonal, latticized and reinserted into the larger matrix in a step-wise fashion.

## Results

*Simulated networks*

A 1000-node lattice with average degree $k$=10 was simulated by creating a 1000×1000 adjacency matrix and assigning links along the 5 sub- and super-diagonals. The lattice was rewired at varying probabilities $p$, ranging from 0 to 1, where $p$=1 resulted in a completely random graph. The rewiring regime used was previously developed by Humphries and Gurney(Humphries and Gurney 2008) and is described in detail in the methods. The clustering coefficient, path length, and $\omega$, were calculated at each rewiring probability (Figure 2a). The small-world regime was defined qualitatively by Watts and Strogatz(Watts and Strogatz 1998) as the region in which clustering (squares) is high and path length (circles) is low. When the plot for $\omega$ is overlaid on the Watts and Strogatz plot, there are three crossing points that are of particular interest. When $\omega$ crosses the zero point, the network is balanced such that path length is as close to random as clustering is to a lattice. This would be the point where the network exhibits ideal small-world properties with a balance between clustering and path length. As the $\omega$ curve moves toward -1 the trade-off between $C$ and $L$ favors $C$. When the $\omega$ curves crosses the $L$ curve, the trade-off is such that the network falls out of the small-world regime and becomes more like a lattice network. Similarly, as the curve moves toward +1, the balance favors $L$ and when the $\omega$ curves crosses the $C$ curve, the network becomes more like a random network. Recall that $\omega$ defines a *continuum* from lattice to small-world to random. No precise cut-off points define small-worldness, but the proximity of $\omega$ to 0 indicates small-world tendencies. However, should one decide that a cut-off is desirable, in this example, the small-world region approximately spans the range $-0.5 \leq \omega \leq 0.5$. The small-world coefficient, $\sigma$, is shown in a similar fashion (Figure 2b). According to the definition of the small-world coefficient, a network is considered small-world if $\sigma > 1$. However, it appears that the network is considered small-world across all rewiring probabilities except for $p=1$. Thus, one might question the meaning of a network that is almost completely rewired to the level of a random network yet has $\sigma > 1$.

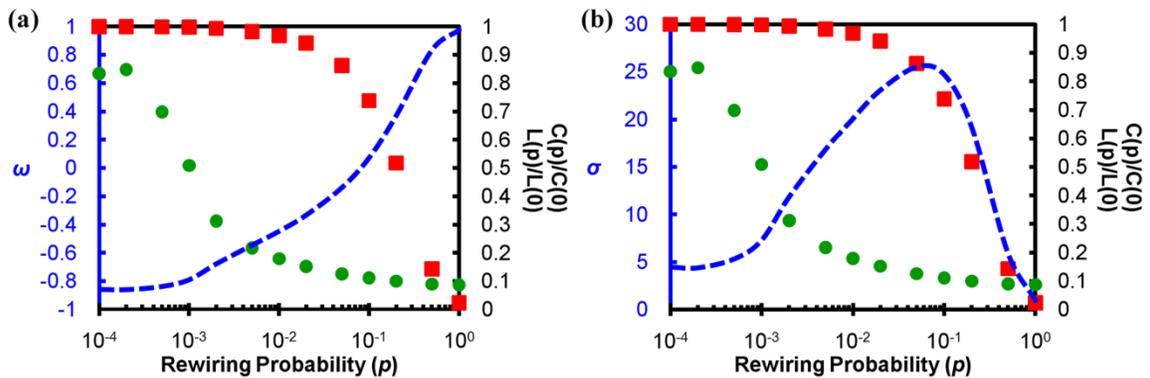

**Figure 2.** Dynamics of $\omega$ and $\sigma$. A 1000-node lattice network was simulated and rewired randomly at varying rewiring probabilities ($p$). (a) Clustering (squares), path length (circles), and the resulting $\omega$ curve (dashed line) is shown. (b) Similarly, the resulting $\sigma$ curve (dashed line) is shown. The axis on the left in both plots indicates the $\omega$ and $\sigma$ values, respectively. The axis on the right in both plots indicates the level of clustering or path length with respect to the $C$ and $L$ of the lattice at $p=0$.

The simulated network analysis shown in Figure 1 was repeated across networks having $N=100$, 500, 1000, 5000, and 10,000 nodes. It is important to note the consistency of $\omega$ in the range $-0.5 \leq \omega \leq 0.8$ across different network sizes (Figure 3a). This suggests that networks with different sizes, yet similar properties, will have similar $\omega$ values, thus making it easier to compare these networks. As stated earlier, values of $\sigma$ trend higher as network size increases, as seen in the peak $\sigma$ value range from 3 to 223 (Figure 3b). In addition, as seen in Figure 2b, networks are classified as small-world across most rewiring probabilities.

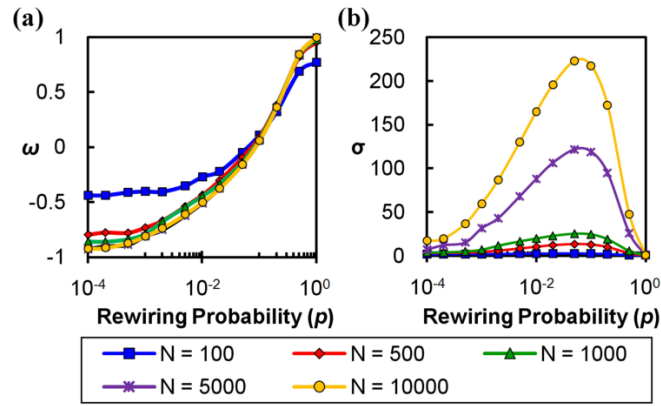

**Figure 3.** Demonstration of validity of $\omega$ by comparing networks of differing sizes. Networks of varying sizes were simulated and rewired over a range of probabilities as discussed previously. For each network, an $\omega$ curve (a) and a $\sigma$ curve (b) are shown. While $\sigma$ is highly dependent on the size of the network, $\omega$ curves demonstrate greater consistency over a range of network sizes. This consistency demonstrates the utility of $\omega$ in making cross-subject or cross-network comparisons.

Note that as the size of the network is reduced, the tails of the $\omega$ curve deviate from the bounds of the [-1,1] interval, particularly from -1. This finding is due to the fact that in a very small network the real path length never becomes very large relative to the path length of the random network. In the equation for $\omega$ a "pure lattice" approaches -1 because the real path length is so much larger than the random path length that the $L_{rand}/L$ of equation (4) approaches zero. If the network is truly small, the path length never exceeds that of the random network by orders of magnitude as observed for large networks. Nevertheless, the middle portion of the curve for all network sizes overlaps. The tails of the $\omega$ curve can also be affected by the density of connections as presented in the supplemental materials (Figure S1). It is important to note that similar curve alterations were observed when evaluating the normalized clustering and path length using the plotting method of Watts and Strogatz (see Figure 1). To account for these deviations from predicted behavior, the evaluation of the small-world regime for

$\omega$ should utilize plots of *C* and *L* for networks of equivalent size and edge density. This method was used to evaluate several real-world networks presented below.

*Real World Networks*

Comparing the $\sigma$ values obtained for several biological, social, and technological networks (Table 1) reveals that most networks evaluated here have $\sigma>1$ and are, therefore, considered small-world. However, is the most highly clustered network, the *C. elegans* metabolic network, less small-world than the e-mail network? Clearly the size of these networks is different, and it is not clear how to take this difference into account when evaluating the small-world properties. Similarly, should one interpret that the karate and protein interactions networks share similar properties because their $\sigma$ values are close to each other? Utilizing the small-world measurement $\omega$ with a small-world criterion defined specifically for each system based on the number of nodes and the average degree, conclusions and classification of these networks are quite different. For example, the e-mail network achieves the largest $\sigma$ value, but the $\omega$ value indicates that it is closer to a random network than a small-world network. Likewise, although the karate and protein interactions networks have $\sigma$ values that are similar, the $\omega$ values indicate that both networks are small-world, but the protein network trends closer to a random topology. It is worth noting that both metrics were able to determine that the word adjacency network is not small-world. However, whereas $\sigma$ can only say the network is not small-world, $\omega$ indicates whether this network is more like a lattice or random network; in this case, the word adjacency network trends toward a random topology.

| Network | N | K | C | $C_{rand}$ | $C_{latt}$ | L | $L_{rand}$ | $\sigma$ | $\omega$ |
|---|---|---|---|---|---|---|---|---|---|
| Karate (Zachary 1977) | 35 | 4.46 | 0.55 | 0.31 | 0.65 | 2.41 | 2.24 | 1.66 | 0.08 |
| Word adjacency (Newman 2006) | 112 | 7.59 | 0.17 | 0.19 | 0.69 | 2.54 | 2.49 | 0.89 | 0.73 |
| Football (Girvan and Newman 2002) | 115 | 10.66 | 0.40 | 0.08 | 0.67 | 2.51 | 2.24 | 4.67 | 0.29 |
| *C. elegans* (metabolic) (Duch and Arenas 2005) | 453 | 8.94 | 0.65 | 0.28 | 0.80 | 2.66 | 2.50 | 2.18 | 0.12 |
| E-Mail (Guimera et al. 2003) | 1133 | 9.62 | 0.22 | 0.09 | 0.55 | 3.60 | 3.27 | 8.14 | 0.56 |
| Protein interactions (Jeong et al. 2001) | 1539 | 2.67 | 0.07 | 0.04 | 0.19 | 6.81 | 5.69 | 1.47 | 0.47 |
| Internet | 22963 | 4.22 | 0.23 | 0.09 | 0.68 | 3.84 | 3.58 | 2.28 | 0.51 |

**Table 1.** Network statistics of several well-known biological, social, and technological networks. Statistics shown include network size (*N*), degree (*K*), clustering coefficient (*C*), clustering of an equivalent random network ($C_{rand}$), clustering of an equivalent lattice network ($C_{latt}$), path length (*L*), and small-worldness metrics ($\sigma$ and $\omega$). The networks were each obtained from previous work as referenced.

Another illustration that demonstrates the advantage of $\omega$ can be seen in the football network (Figure 4). The original network is presented here flanked by an equivalent lattice and random network using the Kamada-Kawai model, an algorithm used to optimize the spatial representation of a graph based on the connections (Kamada and Kawai 1989). The small-world coefficient for each network is $\sigma$=3.49, 4.67, and 0.96 for lattice, real and random network, respectively. Although the real network is classified as small-world, the equivalent lattice network is also categorized as small-world according to $\sigma$. The aberration seen here clearly demonstrates the drawback of comparing the network to an equivalent random network alone. As described earlier, the clustering in a random network, $C_{rand}$, has a great influence on the value of $\gamma$, thus ultimately impacting $\sigma$. This occurs because the magnitude difference between *C* and $C_{rand}$ is often large compared to that of *L* and $L_{rand}$; thus, most lattice networks will be classified as small-world because $\sigma$ will always be greater than 1 unless the network is completely random. Such influence is not seen with $\omega$, where $\omega$ equals -0.60, 0.29, and 0.89, for lattice, real, and random network, respectively. It can be seen here that $\omega$ provides a quantitative measure of the small-world properties over a spectrum of network topologies.

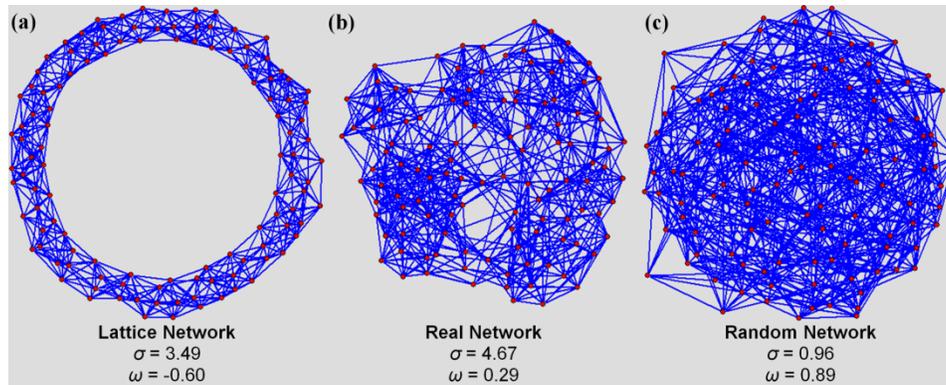

**Figure 4.** Comparison of the football network (b) to its equivalent lattice (a) and random network (c). Each network is represented using the Kamada-Kawai algorithm. Although the network itself is considered small-world ($\sigma$=4.67), the latticized network is also considered small-world ($\sigma$=3.49). Values of $\omega$ were -0.60, 0.29, and 0.89 for the lattice, real, and random network, respectively.

*Brain networks*

In this section the utility of $\omega$ is demonstrated using data from a study comparing the effects of an exercise program in older adults previously completed by our group(Joyce et al. 2010). There was no significant difference for the graph metrics $C$ and $L$ between the control and exercise group (Table 2). However, when the mean small-world coefficient for both groups was compared, they were found to be significantly different. Such a group difference was not seen with $\omega$, which found no significant difference between the groups. From an intuitive standpoint, if clustering in a group of networks were to increase, then the small-world coefficient should also increase. However, despite the groups having similarly sized networks, the $\sigma$ value in the exercise group was lower. Given the similar levels of clustering and path length for both groups, the significant difference seen in $\sigma$ suggests that another factor may influence the calculated value.

|  | $N$ | $C$ | $L$ | $K$ | $\sigma$ | $\omega$ |
|---|---|---|---|---|---|---|
| Control Group | 13488 | 0.32 | 4.14 | 45.18 | 10.28 | 0.28 |
| SD | 327 | 0.02 | 0.14 | 0.46 | 1.22 | 0.05 |
| Exercise Group | 12950 | 0.34 | 4.62 | 46.25 | 5.62 | 0.19 |
| SD | 662 | 0.03 | 0.49 | 0.59 | 2.37 | 0.10 |

**Table 2.** Mean statistics and standard deviations (SD) for real brain data comparing network changes due to an exercise regime. The exercise group showed an increase, but nonsignificant increase for both clustering and path length. An investigation of small-world properties revealed a significant difference for $\sigma$. In contrast, a nonsignificant difference was found for $\omega$, matching the findings for clustering and path length more closely.

Plotting the small-world values against the clustering of the network provides some explanation for the differing results for $\sigma$ and $\omega$ (Figure 5). A comparison of $\sigma$ with network clustering reveals a weak relationship with an $R^2=0.4135$. In contrast, the value for $\sigma$ and the equivalent random network is almost perfect with $R^2=0.9968$. These results suggest that the main factor driving changes in $\sigma$ is the clustering of the random network. In contrast, when $\omega$ is compared to network clustering, its value more accurately reflects clustering in the real networks without undue influence of clustering in the lattice network.

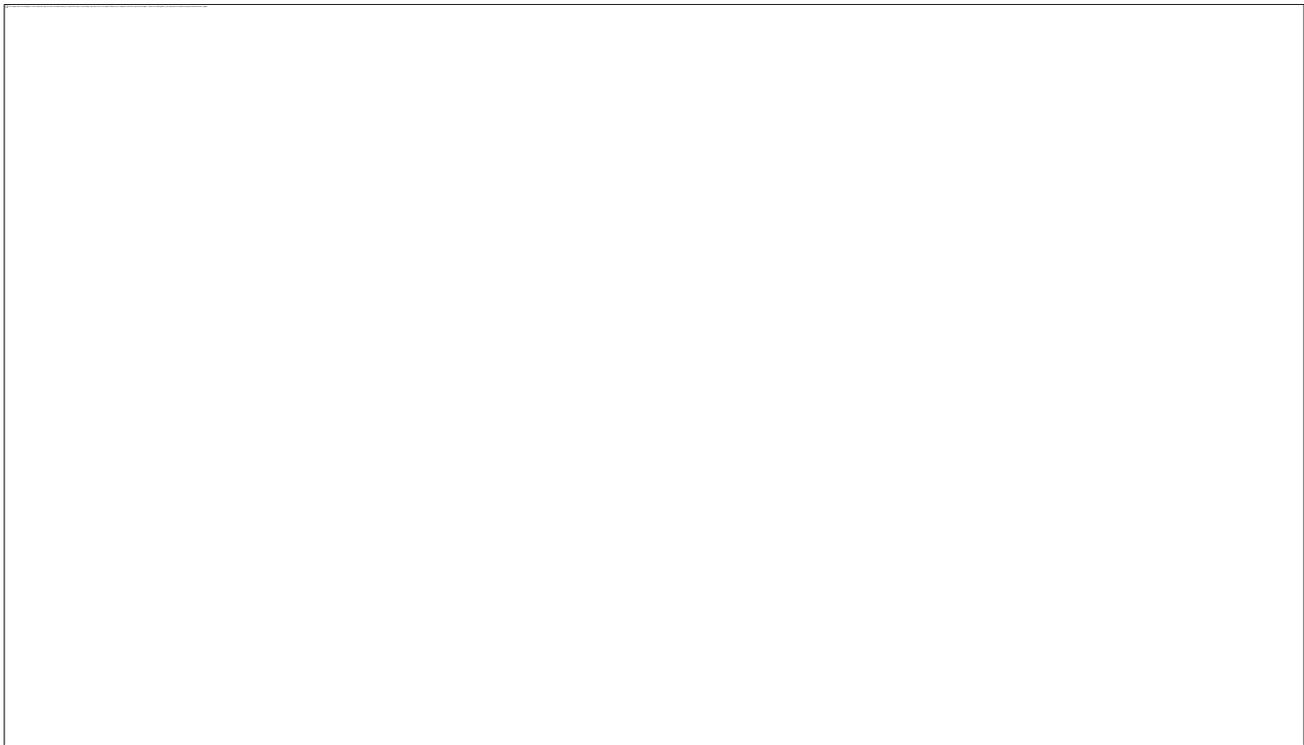

**Figure 5.** Correlations between small-world coefficients and clustering coefficient in brain networks (a) and real-world networks (b). The clustering coefficient of the original networks and clustering of equivalent random network ($C$ and $C_{rand}$, respectively) were compared with $\sigma$. In addition, $\omega$ was compared to $C$ and clustering of the equivalent lattice networks ($C_{latt}$). $R^2$ values indicate goodness of fit of the trendline (solid line). In both groups $\sigma$ shows a stronger association with $C_{rand}$ compared to $C$. In contrast, $\omega$ shows a strong association with $C$ without undue influence by $C_{latt}$.

**Discussion**

We have introduced a small-world measurement, $\omega$, that more accurately quantifies network small-world properties as originally defined by Watts and Strogatz. In our exploration of networks, the traditionally used small-world metric $\sigma$ appears to have high sensitivity in classifying small-world networks. However, it appears that this high sensitivity is coupled with low specificity, resulting in networks being classified as small-world when they are essentially random with only minor clustering present. In fact, the small-world coefficient has been found to classify nearly all networks as small-world ($\sigma>1$) unless the network is a completely random graph. Another drawback of $\sigma$ is that its value gives no sense of where along the spectrum of lattice to random it falls. Greater $\sigma$ values are associated with higher levels of small-worldness; however, this may not be the case as $\sigma$ values do not increase monotonically over varying rewiring probabilities. As seen in Figure 3b, with the exception of the peak, a particular $\sigma$ value can represent networks with radically different topological properties, thus making network comparisons unreliable; $\sigma$ can determine if a network is a random or not, but it cannot effectively determine small-worldness in a network. In our investigation, $\omega$ was found to increase monotonically across all networks used in this study, thus facilitating its use as tool for comparing networks. Unlike the calculation of $\sigma$, the calculation of $\omega$ involves comparing the clustering coefficient of the network to a lattice network, thereby allowing one to determine how similar the original network is to its lattice equivalent. Using this approach provides inherent scaling and determines how much the original network is like its lattice or random equivalents.

*Limitations*

One possible reason why comparisons with network lattices have not been used in the literature up to this point is the length of time it takes to generate lattice networks, particularly for large networks. One appeal of comparing the original network to only a random network is the rather fast processing time to generate the random network. Although latticization is fast in smaller networks, large networks such as functional brain networks and the internet can take several hours to generate and optimize. The latticization procedure in this manuscript utilizes an algorithm developed by Sporns and Zwi in 2004(Sporns and Zwi 2004), but the algorithm was used on much smaller datasets. Based on some relatively minor modifications (see Methods), we were able to use this latticization method on brain networks with >15,000 nodes. As processor speed increases, or perhaps with the development of a more computationally efficient algorithm, it will become possible perform latticization on very large datasets.

Observations of $\omega$ show that as network size decreases, the range of $\omega$ tends to decrease. As seen in the 100-node network in Figure 2, the minimum $\omega$ value is greater than -1. This occurs because in smaller networks, the equivalent lattice tends to have shorter path length. Since the path length in the lattice network is closer to that of the random network, the $L_{rand}/L$ term in Equation 4 deviates from 0; thus, $\omega$ does not approach -1. $\omega$ is also limited by networks that have very low clustering that cannot be appreciably increased, such as networks with "super hubs" or hierarchical networks. In hierarchical networks, the nodes are often configured in branches that contain little to no clustering. In networks with "super hubs," the network may contain a hub that has a node with a degree that is several times in magnitude greater than the next most connected hub. In both these networks, there are fewer configurations to increase the clustering of the network. Moreover, in a targeted assault of these networks, the topology is easily destroyed(Albert et al. 2000). Such vulnerability to attack signifies a network that may not be small-world. However, due to the normalization of clustering in Equation 4, the closeness of $C$ and $C_{latt}$ can introduce a bias that makes the network appear as if it has high clustering.Similar to the WS model $\omega$ may become unreliable for very sparse networks, thus it is crucial to examine properties like edge density to determine if $\omega$ is applicable to the network of interest. In our investigation, one network that may fall into this category is the protein interactions network due to its low edge density and 60% of its edges forming trees. Thus, one should use caution when trying to attribute small-world properties to such networks.

The Watts and Strogatz model is based on a regular lattice with nodes of equal degree. Observed in it are $C$ and $L$ curves in relation to a specific network size and degree distribution. Changing the number of nodes and degree distribution will certainly change these curves, thus affecting the extent of the small-world criterion. For this manuscript, the small-world range was defined to be [-0.5,0.5], but this interval may not be static across differently sized networks. In an ideal network, $\omega$ ranges from -1 to 1; a value of 0 indicates a network that is in perfect balance between high clustering and low path length, while the extremes at -1 and 1 represent a pure lattice and random network, respectively. The values at ±0.5 represent the maximum values before the networks exhibit properties that are more like a random network or more like a lattice network. For example, a network with an $\omega$ of 0.6 more closely resembles a random network than a small-world network. This region of small-worldness [-0.5,0.5] is very sensitive to detecting small-world networks. One could be more stringent and require the $\omega$ value to be closer to zero to increase specificity. The cut-off to be used depends on the goals of the study with values closer to zero will be more specific while values closer to 0.5 will be more sensitive. It is important to realize that this range is very sensitive and provides a guideline for classifying small-worldness.

Future investigation of network topology, particularly in relation to extreme hubs and the heterogeneity of node degree, is warranted. However, such exploration is beyond the scope of this work. The intention of $\omega$ is to indicate the level of small-worldness along a continuum and whether a network exhibits more lattice-like or more random-like behavior. It is worth noting that $\omega$ acts as a summary statistic and estimates the overall small-world properties in a network. Mean graph metrics and summary statistics should not serve as an endpoint for network analysis. Instead, these metrics should be used in conjunction with other analyses that can provide further information about the complex organization of the network. Nonetheless, $\omega$ is an appealing metric because networks with similar properties share the same value regardless of network size, and in most cases, researchers compare networks of similar size and topology, as seen with brain imaging data. In this regard, $\omega$ may be a better choice as it is sensitive to changes in these networks.

**Conclusion**

In this manuscript we have introduced a new network metric, $\omega$, and have shown that small-world networks may be less ubiquitous than suggested in the current literature. We have demonstrated that this metric more accurately identifies small-world networks and it is able to determine if a network has more lattice- or random-like properties. In addition, this metric is less sensitive to the size of a network and benefits from inherent scaling, which provides a powerful tool for comparing and ranking small-world properties in various systems. However, the strength of this metric truly lies in its ability to compare similarly sized networks as it allows for a more direct comparison of network properties. The capacity to describe a network as more random- or lattice-like provides considerable benefits to studies investigating dynamic changes in a network. Moreover, direct comparison of small-world properties is useful for group studies, such as brain networks, where understanding the network topology in a particular population may provide further insight into a disease or pathology. This metric provides a useful tool for studying complex systems, but more importantly, it truly characterizes where a network falls in the Watts and Strogatz small-world model.

**Acknowledgements**

**Supporting Information**

**Effect of network density on $\omega$**

The main text discusses the effect of network size on the form of the ω curve. Similar simulations were performed to examine the effects of varying the network density. To this end, four 1000-node regular lattice networks were simulated at varying densities (density = 1%, 5%, 10%, and 20%). Each network was rewired at random according to the Watts and Strogatz rewiring regime(Watts and Strogatz 1998). The resulting clustering coefficient, path length, and ω values are shown versus each corresponding rewire probability (Figure S1).

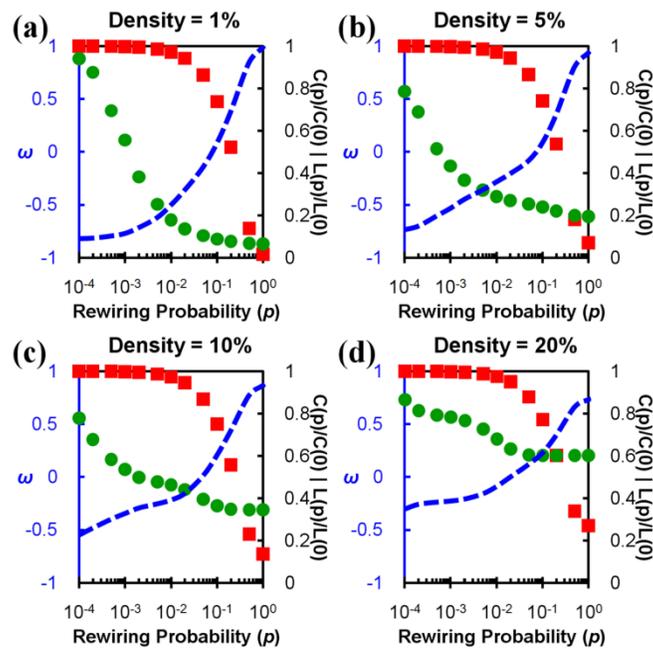

**Figure S1.** Effect of network density on $\omega$. Four 1000-node networks were simulated at densities of 1%, 5%, 10%, and 20%. Each network was rewired at a given rewire probability to span the range from lattice to small-world to random. Clustering and path length as well as $\omega$ values for each network are shown.

In general, as the density of a network increases, the nodes of the network become more connected. Therefore the average path length in dense networks is generally shorter. Thus, the maximum improvement in path length due to random rewiring becomes limited for these dense networks, and ω of the lattice network cannot achieve very negative values. This provides an explanation for the curve seen in Figure 2 for a network size of 100, where the lower tail of the omega curve does not achieve -1. This small network has an average degree of 16, and therefore a density of approximately 32%.

**Real World Networks**

Small-world properties were investigated in ten networks found in the literature and various databases. These networks include various biological, social and technological networks. As seen in Table S1, the small-world coefficient, $\sigma$, indicates that most of these networks are small-world. In contrast, using $\omega$, the classification of some of these networks as small-world comes into question. Moreover, $\omega$ allows for ranking of these networks on a continuum between lattice and random. As shown in Figure 5, $\sigma$ does not characterize clustering in networks and is greatly influenced by clustering in the random network. These results parallel the results found in brain network data (Figure 4).

| Network | $N$ | $K$ | $C$ | $C_{rand}$ | $C_{latt}$ | $L$ | $L_{rand}$ | $\sigma$ | $\omega$ |
|---|---|---|---|---|---|---|---|---|---|
| Karate (Zachary 1977) | 35 | 4.46 | 0.55 | 0.31 | 0.65 | 2.41 | 2.24 | 1.66 | 0.08 |
| Dolphin (Lusseau et al. 2003) | 62 | 5.13 | 0.26 | 0.10 | 0.57 | 3.36 | 2.71 | 2.03 | 0.35 |
| Word adjacency (Newman 2006) | 112 | 7.59 | 0.17 | 0.19 | 0.69 | 2.54 | 2.49 | 0.89 | 0.73 |
| Football (Girvan and Newman 2002) | 115 | 10.66 | 0.40 | 0.08 | 0.67 | 2.51 | 2.24 | 4.67 | 0.29 |
| Jazz (Gleiser and Danon 2003) | 198 | 27.70 | 0.62 | 0.26 | 0.76 | 2.24 | 1.99 | 2.08 | 0.08 |
| US airlines (1997) (Batagelj and Mrvar 2006) | 332 | 12.81 | 0.63 | 0.43 | 0.73 | 2.74 | 2.48 | 1.32 | 0.04 |
| *C. elegans* (metabolic) (Duch and Arenas 2005) | 453 | 8.94 | 0.65 | 0.28 | 0.80 | 2.66 | 2.50 | 2.18 | 0.12 |
| E-Mail (Guimera et al. 2003) | 1133 | 9.62 | 0.22 | 0.09 | 0.55 | 3.60 | 3.27 | 8.14 | 0.56 |
| Protein interactions (Jeong et al. 2001) | 1539 | 2.67 | 0.07 | 0.04 | 0.19 | 6.81 | 5.69 | 1.47 | 0.47 |
| Internet | 22963 | 4.22 | 0.23 | 0.09 | 0.68 | 3.84 | 3.58 | 2.28 | 0.51 |

**Table S1.** Network statistics of several well-known biological, social, and technological networks. Statistics shown include network size ($N$), degree ($K$), clustering coefficient ($C$), clustering of an equivalent random network ($C_{rand}$), clustering of an equivalent lattice network ($C_{latt}$), path length ($L$), and small-worldness metrics ($\sigma$ and $\omega$). The networks were each obtained from previous work as referenced.

**Network latticization**

Lattice networks were generated using a modified version of the Sporns-Zwi "latticization" algorithm(Sporns and Zwi 2004)found in the brain connectivity toolbox (http://www.brain-connectivity-toolbox.net)(Rubinov and Sporns 2010). The procedure is based on a Markov-chain algorithm that maintains node degree and swaps edges with uniform probability; however, swaps are only carried out if the resulting matrix has entries that are closer to the main diagonal. The algorithm requires two inputs: the adjacency matrix of the network and the number of iterations (i.e., the number of times each node is rewired on average). Using the algorithm in its present form generates a network that places most of the nodes along the main diagonal of the matrix. However, because this algorithm is based on random swaps, the clustering in the network is not necessarily maximized. Increasing the number of iterations may increase clustering, but this often comes at the cost of increased processing time. Although this algorithm works well with smaller networks, it becomes computationally burdensome with larger networks like the internet, which takes several hours to process.

To address these issues, a new latticization algorithm was developed that slightly modifies the Sporns- Zwi algorithm. The latticization algorithm used in this study utilizes a two-step procedure to produce a lattice network with optimized clustering (see Figure S2 for schematic). The first step simply uses the Sporns-Zwi algorithm with five (5) iterations on the original adjacency and produces an input matrix, $A_{ij}(in)$ (i.e., the matrix that is used by the clustering optimization algorithm). Five iterations are used because it appears to strike a balance between processing time and networks with most nodes along the main diagonal. The second step uses the input matrix, $A_{ij}(in)$, and a user-defined number of repetitions, $R$, to ultimately generate a lattice network, $A_{ij}(latt)$. The clustering optimization algorithm uses the Sporns-Zwi algorithm, but uses one (1) iteration, to produce an output matrix, $A_{ij}(opt)$. Following this procedure, the clustering of $A_{ij}(opt)$ is compared to $A_{ij}(in)$. If $A_{ij}(opt)$ has higher clustering, it replaces $A_{ij}(in)$ as the new input matrix. If $A_{ij}(opt)$ has lower clustering, then $A_{ij}(in)$ is kept as the input matrix. Using the old or new input matrix, the clustering optimization procedure is performed until the number of user-defined repetitions is fulfilled. The resultant matrix, $A_{ij}(latt)$, is a lattice network with optimized clustering. This latticization algorithm is useful for small and mid-sized networks as it ensures highly clustered networks. Furthermore, this algorithm takes less processing time than simply increasing the number of iterations in the Sporns-Zwi algorithm, which does not necessarily produce optimized clustering.

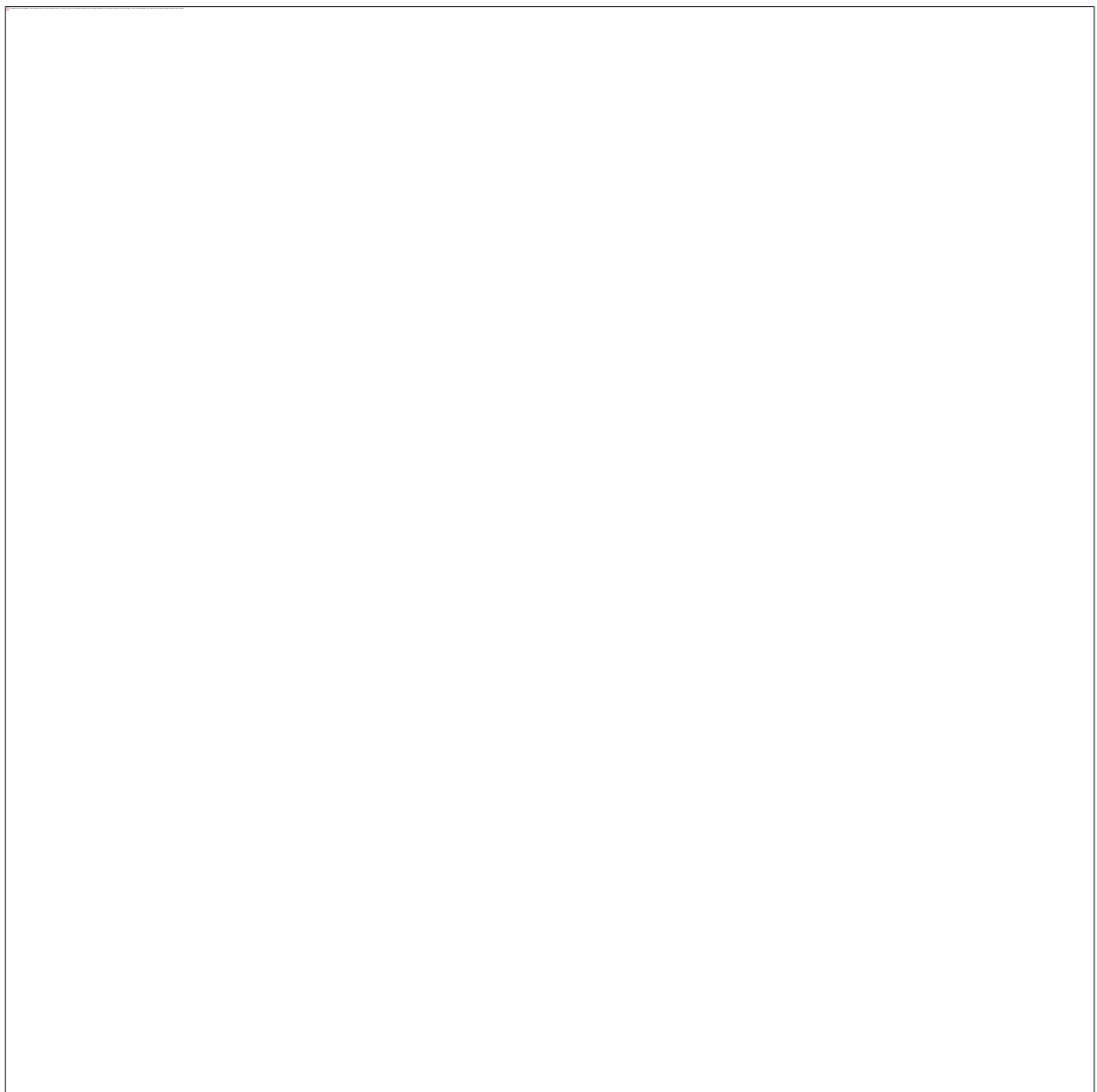

**Figure S2.** Schematic of network latticization procedure

As mentioned earlier, larger networks become a computational burden for the Sporns-Zwi algorithm as well as the latticization algorithm used in this study. To address this, an alternative method was developed for larger networks like the internet or voxel-based functional brain networks. Instead of using the latticization algorithm on the entire network, the algorithm is used on smaller subnetworks (or partitions) along the main diagonal. For example, instead of using the latticization algorithm on a full 5000×5000 matrix, smaller 500×500 partitions along the main diagonal can be latticized. Using this approach decreases processing time because smaller networks are used by the algorithm.

The "sliding window" algorithm detailed here was used on all the larger networks in this study (> 1000 nodes). As seen in Figure S3, the algorithm uses a three-step procedure to generate a lattice network. The initial step still uses the Sporns-Zwi algorithm with five (5) iterations. This step was used for all matrices regardless of network size as it provides a good starting point for the latticization algorithm. The second and third step use the input matrix, $A_{ij}(in)$, a user-defined number of repetitions, $R$, and a window size, which represents the size of the partitions to be latticized.

Before using the sliding window procedure, corner latticization is performed. Assuming a ring-like structure to the network, nodes toward the beginning and end of the main diagonal are considered close to each other. Connections among these nodes populate the corners of the matrix, thus the sliding window procedure bypasses these nodes. Based on the input window size, the corner latticization step extracts four equally sized corners from the input matrix and joins them to form a smaller "corner matrix." Using the latticization algorithm detailed in Figure S2, the corner matrix is latticized. After latticization, the corner matrix is split into four part and reinserted into the input matrix, replacing the entries of the original matrix. Using this new matrix, the sliding window procedure is used on the full input matrix. Along the main diagonal, a partition equal to the input window size is extracted, latticized, and reinserted into the matrix. Moving half the distance of the window along the main diagonal, another partition is extracted and latticized as before. The sliding window procedure continues until the entire matrix is covered resulting in a lattice network.

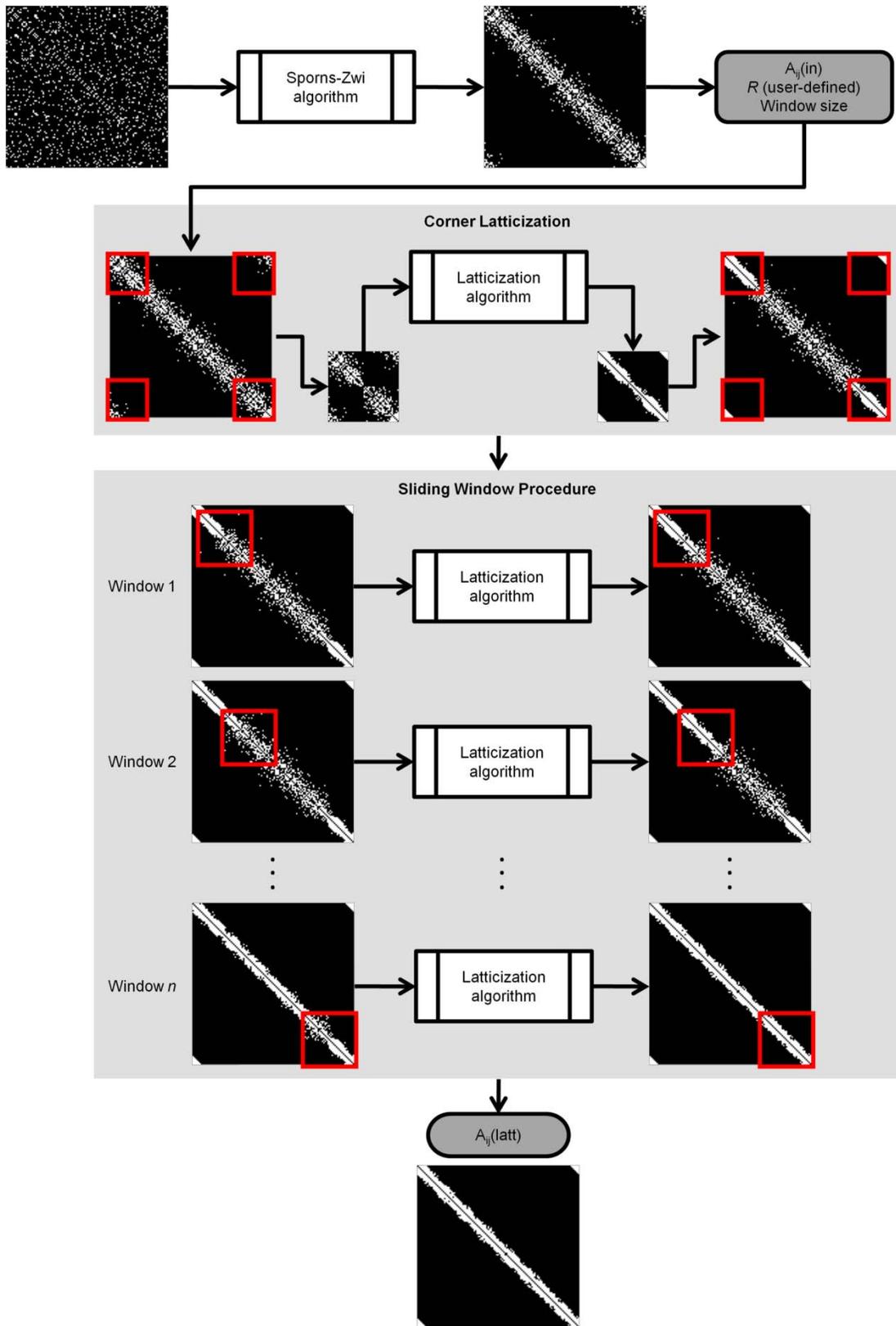

**Figure S3.** Schematic of latticization procedure for larger networks. In larger networks, corner latticization and the sliding window procedures are used to speed up processing time.

**Network Resources**

Biological, social, and technological networks were obtained from various sources. All networks in this study were analyzed as unweighted and undirected graphs. For disconnected graphs, network analysis was done on the largest component of the network. The US airlines network(Batagelj and Mrvar 2006)was obtained from the Pajek datasets (http://vlado.fmf.uni-lj.si/pub/networks/data/). The e-mail(Guimerà et al. 2003), jazz(Gleiser and Danon 2003), and *C. elegans*metabolic network(Duch and Arenas 2005)were obtained from Alex Arenas' network datasets (http://deim.urv.cat/~aarenas/data/welcome.htm). The karate(Zachary 1977), word adjacency(Newman 2006), football(Girvan and Newman 2002), dolphin(Lusseau et al. 2003), and internet networks were obtained from Mark Newman's network data sets (http://www-personal.umich.edu/~mejn/netdata/). The internet network is from unpublished data by the University of Oregon Route Views Project (http://routeviews.org/).

**Brain imaging data collection**

*Scanning protocol*

Brain imaging data was collected from 11 healthy older adults as part of a separate study evaluating an exercise program(Joyce et al. 2010). All data reported here are from post-treatment scans with subjects either in the control or treatment group. The scanning protocol can be found in supplemental materials. Functional MRI (fMRI) data was collected on a 1.5 T GE twin-speed LX scanner with a birdcage head coil (GE Medical Systems, Milwaukee, WI). Functional imaging was performed using multi-slice gradient-EPI images (TR/TE=2000/40ms; field of view=24cm (frequency) × 15cm (phase); matrix size=96×86, 40 slices, 5 mm thickness, no skip; voxel resolution=3.75mm×3.75mm×5mm). Subjects performed no task, but were asked to keep their eyes open for the 6 minute 20 second (190 images) resting fMRI scan. Images were motion corrected, normalized to MNI (Montreal Neurological Institute) space, and resliced to a 4×4×5mm voxel size using SPM99 (Wellcome Trust Centre for Neuroimaging, London, UK). All participants gave written informed consent approved by the Institutional Review Board at Wake Forest University School of Medicine, Winston-Salem, NC.

*Network analysis*

Whole-brain functional connectivity was evaluated using graph theory methods on a voxel-by-voxel basis. fMRI time courses were extracted for each voxel in gray matter (approximately 15,000) and band-pass filtered to remove signal outside the range of 0.009–0.08 Hz(Fox et al. 2005; van den Heuvel et al. 2008). Network analysis was based on subject specific gray matter tissue maps with mean white matter and CSF signal regressed from the filtered time series to account for physiological noise. The six rigid-body motion parameters from the motion correction process were also regressed from the time series. A correlation matrix was produced by computing the Pearson correlation between all voxel pairs within fMRI time series. A threshold was applied to the correlation matrix, whereby voxel pairs above the threshold were considered connected and assigned a value of 1, and voxel pairs below the threshold were considered not connected and assigned a value of 0. This binary matrix produces an undirected, unweighted adjacency matrix ($A_{ij}$) representing the whole-brain functional connectivity for each subject. In order to make networks across subjects comparable, the threshold was defined such that the relationship between the number of nodes $N$ and the average node degree $k$ was the same across all subjects. Specifically, the threshold was defined so that $S=\log(N)/\log(k)$ with $S=2.5$ across all subjects for the study(Stam et al. 2007; Supekar et al. 2008). From the adjacency matrix, the following graph metrics were calculated at each node as well as averaged to yield means for the entire network: degree ($k$), clustering coefficient ($C$) and minimum path length ($L$). Details on calculating these graph metrics for an undirected, unweighted graph can be found in Rubinov and Sporns(Rubinov and Sporns 2010).